\documentclass{llncs}
\pdfoutput=1
\usepackage[absolute,overlay]{textpos}
\usepackage{makeidx}  

\usepackage[utf8]{inputenc}

\usepackage{booktabs} 
\usepackage{multirow}

\usepackage{caption}
\usepackage{floatrow}
\usepackage{array}
\usepackage{datetime}
\usepackage{graphicx}
\usepackage{color}
\usepackage{graphbox} 

\usepackage{amssymb} 
\usepackage{enumitem} 

\definecolor{colxxx}{rgb}{.8, .3, 0.0}

\newcommand*{\Dkw}{T20k}
\newcommand*{\Df}{T60k}
\newcommand*{\Dxl}{T400k}

\begin{document}
\begin{textblock*}{10cm}(6cm,2cm)
This is an authors' manuscript version of a paper accepted for proceedings of \emph{TPDL-2018, Porto, Portugal, Sept 10-13}. \\
The final authenticated publication is available online at\\
https://doi.org/will be added as soon as available
\end{textblock*}
\pagestyle{headings}  
\mainmatter              

\title{Content-Based Quality Estimation for Automatic Subject Indexing of Short Texts under Precision and Recall Constraints
}

\author{Martin Toepfer\inst{1} \and Christin Seifert\inst{2}}
\institute{ZBW -- Leibniz Information Centre for Economics, Kiel, Germany\\ 
\email{m.toepfer@zbw.eu}\\ 
\and
University of Twente, Enschede, The Netherlands\\
\email{c.seifert@utwente.nl}
}

\maketitle              

\begin{abstract}
Semantic annotations have to satisfy quality constraints to be useful for digital libraries, which is particularly challenging on large and diverse datasets. Confidence scores of multi-label classification methods typically refer only to the relevance of particular subjects, disregarding indicators of insufficient content representation at the document-level. Therefore, we propose a novel approach that detects documents rather than concepts where quality criteria are met. Our approach uses a deep, multi-layered regression architecture, which comprises a variety of content-based indicators. We evaluated multiple configurations using text collections from law and economics, where the available content is restricted to very short texts. Notably, we demonstrate that the proposed quality estimation technique can determine subsets of the previously unseen data where considerable gains in document-level recall can be achieved,  while upholding precision at the same time. Hence, the approach effectively performs a filtering that ensures high data quality standards in operative information retrieval systems.
\end{abstract}

\section{Introduction}
\begin{figure}[t]
\begin{floatrow}
\ffigbox{ \centering
\includegraphics[width=.98\linewidth]{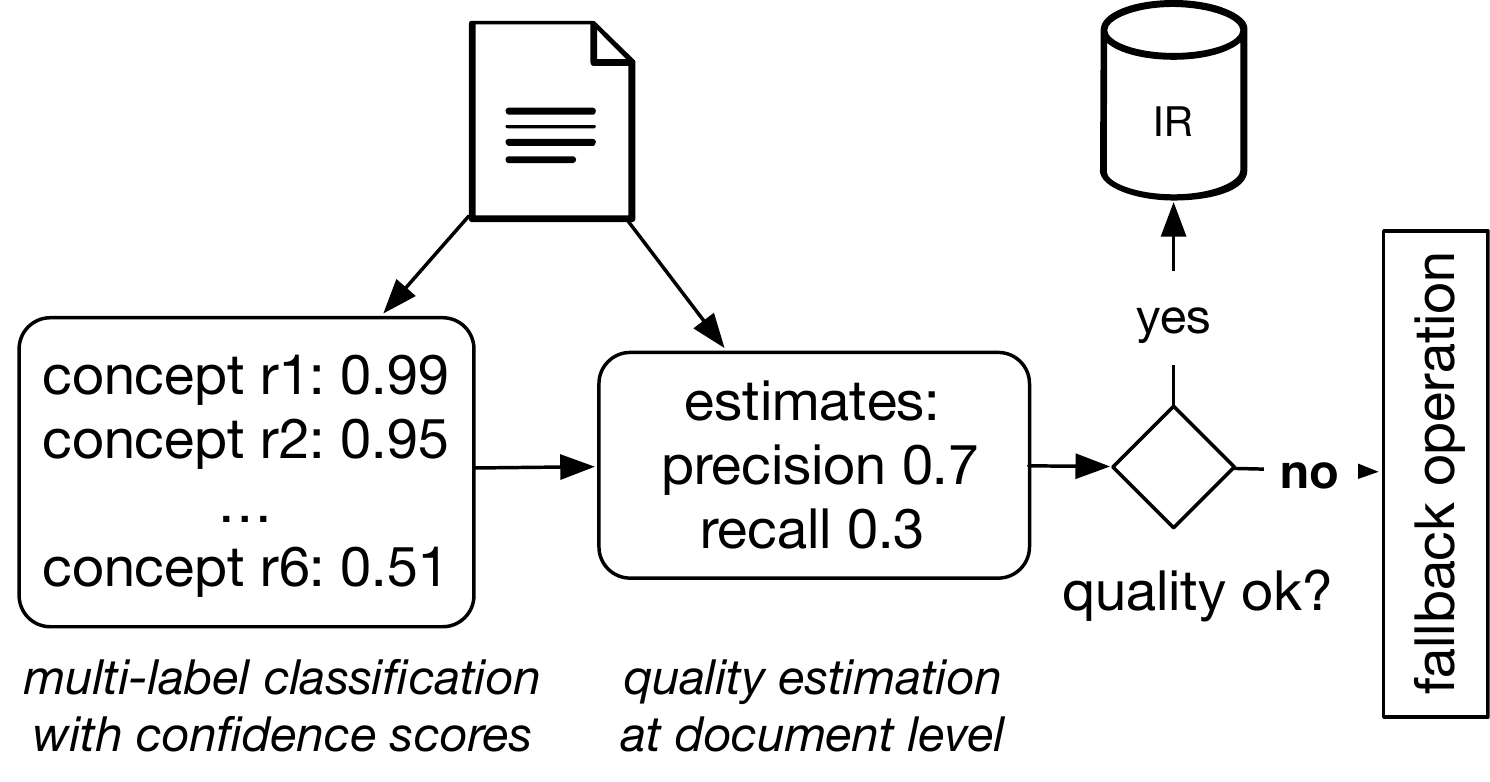}
}{\caption{Schematic overview of the main application context. 
Document-level quality estimation enables filtering of automatic subject indexing results.}
\label{fig:qe4dummies}}
\ffigbox[\Xhsize]{ \centering
\includegraphics[width=\linewidth]{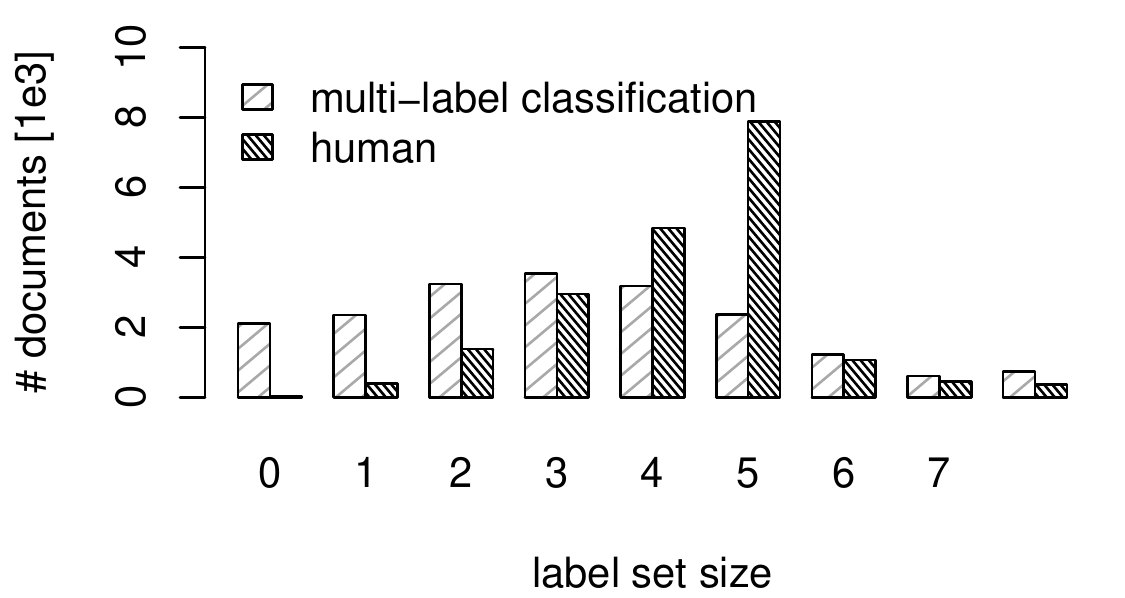}}{\caption{
Illustration of low document-level recall by comparing distributions of label set size (human vs. multi-label classification)
[Dataset: EURLEX].
}
\label{fig:motivation:distribution:mlc}}
\end{floatrow}
\end{figure}

Semantic annotations from automatic subject indexing can improve information retrieval (IR) by query expansion, however, classification performance is a critical factor to gain the benefits~\cite{Trieschnigg2009}.
The relevance of multi-label text classification engendered research in several disciplines. Although considerable progress has been made over the last decades~\cite{Sebastiani2002,Bennett2002,Medelyan2006,LozaMencia2010,Huang2011,Wilbur2014,Toepfer2017,Liu2017},
several challenges remain.
Just to give an example, precision@5 = 52\%~\cite{Liu2017} has recently been reported for a dataset in the legal domain (EURLEX~\cite{LozaMencia2010}), which means that on average per document only half of the five top-ranked subjects matched human annotations. 
Institutional quality requirements, like for instance at libraries, often put severe constraints on precision~\cite{Bennett2017} as well as recall.
It is therefore not sufficient to just apply state-of-the-art algorithms with respect to averaged $f_1$ scores, but further necessary to automatically separate the wheat from the chaff. On large datasets, human quality assessment can be too expensive.
While specific confidence estimation approaches have been proposed for domains like information extraction~\cite{Culotta2004},
automatic subject indexing and multi-label text classification miss essential research in this direction.
We aim to fill this gap since quality estimation is becoming increasingly important for digital libraries to integrate autonomous processes into operative IR systems.
As depicted in Fig.~\ref{fig:qe4dummies}, document-level quality estimates allow to implement filters at the interface to databases which are used for IR.

%
%
%
Most automatic subject indexing methods provide a score for each concept~\cite{Sebastiani2002,Medelyan2008,Huang2011,Wilbur2014},
hence allowing to exclude individual predictions that might be incorrect.
Such precision-oriented filtering removes single label assignments from documents,
leading to lower document-level recall, as exemplified in Fig.~\ref{fig:motivation:distribution:mlc}.
As a direct consequence, 
it becomes difficult to assess document-level quality. As can be seen, the plain number of assigned concepts to a document is not a satisfying indicator, since human indexers\footnote{For brevity, the remainder of this paper will simply uses the terms indexing, indexer, \ldots to refer to subject indexing, subject indexer, \ldots, respectively.} use a wide range of label set sizes.
Interestingly, Sect.~\ref{sec:qe} will point out that uncertainty in recall is an inherent and inevitable phenomenon of multi-label text classification when only a few preconditions are met. 
Thus we conclude that 
concept-level confidence scores must be complemented with document-level estimates, as investigated in this paper.

In summary, the contributions of this work are the following:
\begin{itemize}[topsep=1pt, partopsep=1pt]
  \item We provide a conceptual analysis of confidence and quality estimation for automatic subject indexing.
  \item We propose a quality estimation approach, termed \emph{Qualle}, which combines multiple content-based features in a multi-layered regression architecture. 
  \item We show the impact of different feature groups and the effectiveness of \emph{Qualle} for quality estimation and filtering in an empirical study. 
\end{itemize}

The remainder of the paper starts with a discussion of related work before the central section introduces the quality estimation approach (Sect.~\ref{sec:qe}) followed by experimental results (Sect.~\ref{sec:experiments}).

\section{Related Work}
\label{sec:related-work}
\emph{Confidence scores} are an integral part of many machine learning (ML) approaches for multi-label text classification~\cite{Sebastiani2002,Gibaja2015}. 
For instance, rule-learning typically computes a confidence score for each rule, dividing the number of times the rule correctly infers a class label by the number of times the rule matches in total.
Naive-Bayes approaches use Bayes' Rule to derive conditional probabilities. 
Flexible techniques have been developed to perform probability calibration~\cite{Zadrozny2002}. 
Thus, systems using multi-label classification (MLC) machine learning methods for subject indexing often provide confidence scores for each subject heading. \cite{Sebastiani2002,Gibaja2015}. 
Medelyan and Witten~\cite{Medelyan2008} used decision trees to compute confidence scores for dictionary matches. 
Huang et al.~\cite{Huang2011} similarly applied a learning-to-rank approach on MeSH term recommendation based on candidates from k-nearest-neighbors. 
In general, binary relevance (BR) approaches also provide probabilities for each concept, for instance by application of probability calibration techniques (e.g.~\cite{Wilbur2014}). 
%
Tang et al.~\cite{Tang2009} proposed a BR system which additionally creates a distinct model to determine the number of relevant concepts per document.
In summary, the scores provided by the above mentioned systems are limited to concept-level confidence, that is, referring to distinct subjects.

In the context of classifier combination,
Bennett et al.~\cite{Bennett2002} proposed \emph{reliability-indicator variables} 
for model selection. 
They
identified four types of indicator variables and showed their utility. 
In contrast to their work, we focus on different objectives. 
We apply such features (reliability indicators) for quality estimation, which
in particular comprises estimation of recall. 
By contrast,
\emph{precision-constrained} situations have recently been studied by Bennett et al.~\cite{Bennett2017}. 
%
\emph{Confidence in predictions and classifiers} has recently gained attention in the context of \emph{transparent machine learning} (e.g.~\cite{Ribeiro2016}). 
Contrary to transparent machine learning,
quality estimation does not  
aim to improve interpretability, 
and it thus may be realized by black box ML models.
Nevertheless, quality estimates may be relevant for humans to gain trust in ML.

%
Confidence estimation has been studied in different \emph{application domains}, and it has been noted that different levels of confidence scores are relevant.
For instance, Culotta and McCallum~\cite{Culotta2004} distinguished between field confidence and record confidence (entire record is labeled correctly) in information extraction.
They compared different scoring methods and also trained a classifier that discriminates correct and incorrect instances for fields and records, respectively.

\section{Quality Estimation}
\label{sec:qe}
Our approach to quality estimation (Sect.~\ref{sec:qe:qualle}) stems from an analysis of common practice, as described in the following.

\subsection{Analysis}
\label{sec:qe:analysis}
In the past, quality of automatic subject indexing has been assessed in different ways that have individual drawbacks.
Traditionally, library and information scientists regarded indexing quality, effectiveness, and consistency~\cite{Rolling1981}. Quality assessment that requires human judgements is, however, costly, which can be a severe issue on large and diverse datasets. For this reason, evaluations of automatic subject indexing often just rely on consistency with singly annotated human indexing, yielding metrics which are known as precision and recall. 
As described in Sect.~\ref{sec:related-work}, 
common indexing approaches provide 
confidence scores for each class, denoting posterior probabilities
\( p(y_j = 1 | \mathcal{D} ) \), where $y_j$ refers to a single concept of the controlled vocabulary, 
thus they are referred to as \emph{concept-level confidence} in this work.
Statistical associative approaches derive confidence scores based on dependencies between terms and class labels from examples. 
As a consequence, the performance of these methods largely depends on the availability of appropriate training examples and the stability of term and concept distributions, whereas lexical methods require vocabularies that exhaustively cover the domain.
When concept drift occurs, that is, if observed terms and the set of relevant concepts differ between training data and new data, both types of indexing approaches considerably decrease in performance~\cite{Toepfer2017}. 
Interestingly, since these algorithms merely learn to assign recognized subjects of the controlled vocabulary, they will silently miss to assign relevant subjects not covered by the controlled vocabulary, and moreover they are unable to recognize and represent the loss in document-level content representation.
It is further plausible that these issues are more pronounced when only titles of documents are processed, since for title-based indexing the complete subject content is compressed into only a few words which makes understanding of each single word more crucial compared to processing full texts.
As the evolution of terms and concepts is an inherent property of language~(cf. e.g.~\cite{Tahmasebi2017}), accurate recognition of insufficient exhaustivity is essential in the long term. 
It must be assumed that uncertainty in recall is an inherent and inevitable phenomenon of automatic indexing and multi-label text classification in general.
For these reasons,
in order to guarantee quality, indexing systems must 
gain knowledge relating to classifier reliability based on additional representations (cf.~\cite{Bennett2002}), exploiting information such as out-of-vocabulary term occurrences and document length, just to give an example.
Therefore, instead of concept-level confidence, we propose to address document-level quality directly.

\subsection{Qualle: Content-Based Quality Estimation}
\label{sec:qe:qualle}
\begin{figure}[tpb]
\centering
\includegraphics[width=.7\linewidth]{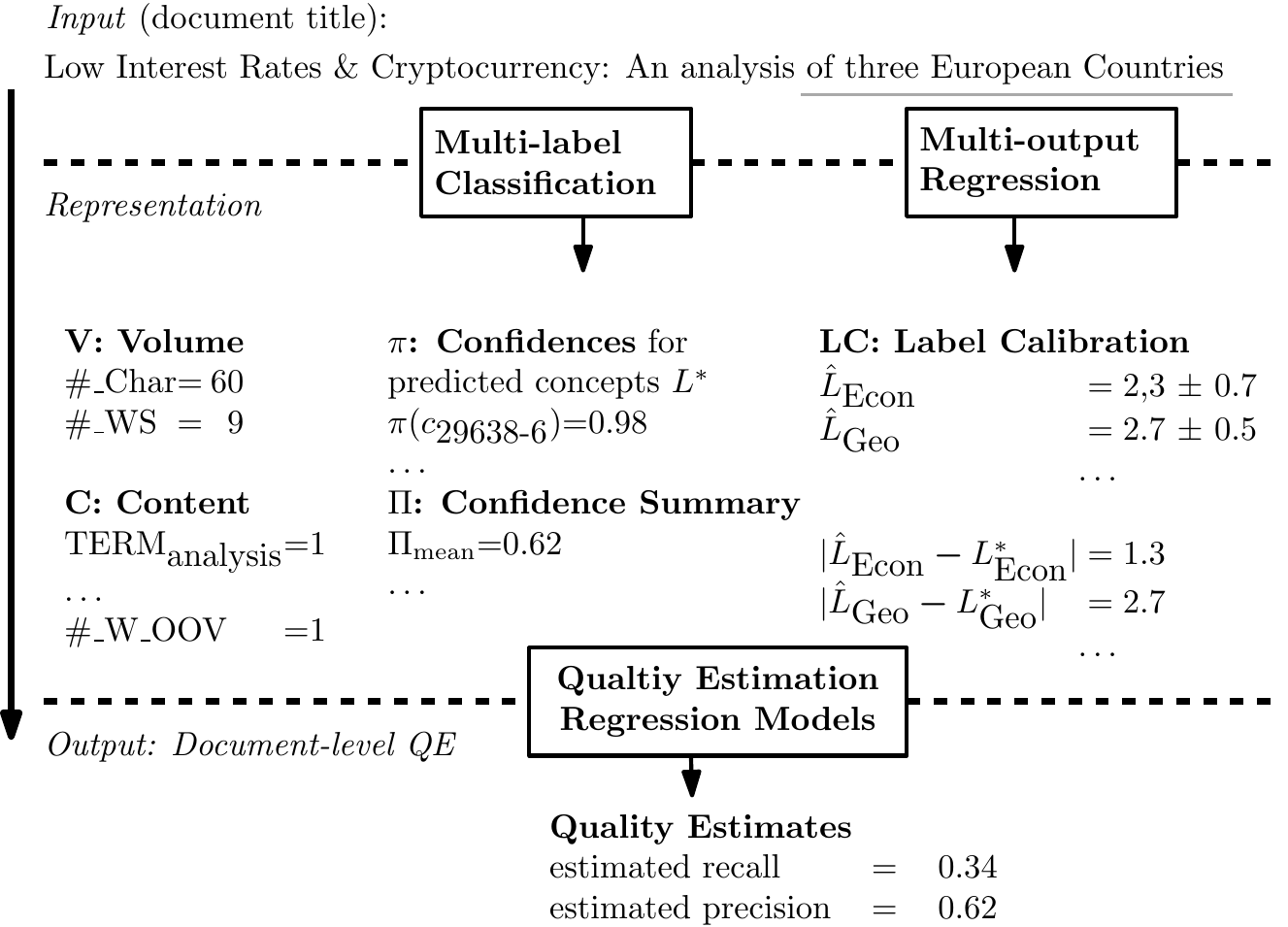}
\caption{
Multi-layered regression architecture
for quality estimation (example).
}
\label{fig:qepred}
\end{figure}

Multi-label classification methods can be tuned by regularization and configuration of thresholds to satisfy constraints on precision.
Hence, the main challenge for our approach on quality estimation, \emph{Qualle}, is to estimate document-level recall.
As a solution,
we propose the architecture which is exemplified in Fig.~\ref{fig:qepred}.
The input layer shows a fictitious title of a document to be indexed, 
which is then represented by multiple features. 
The content is processed by a multi-label text classification method, 
producing a set of concepts and corresponding confidence values, e.g., $\pi (c_{\mathrm{29638-6}}) = 0.98$.\footnote{The concept identifier 29638-6 refers to the concept ``Low-interest-rate policy''.} 
Moreover, a multi-output regression module offers expectations regarding the proper number of concepts for a document (Label Calibration), if possible considering distinct semantic categories, for instance geographic names ($\hat{L}_{\mathrm{Geo}}$) or economics subject terms ($\hat{L}_{\mathrm{Econ}}$), commodities, and much more. 
In the given example, the phrase ``three European countries'' clearly points out that it would be reasonable to assume three geographic names when access to the full text is possible, however, without particularly specifying which ones to choose. The input is not precise enough.
Drawing connections between the predicted concept set $L^*$ and the estimated numbers of concepts $\hat{L}$ can indicate such recall issues.
For instance, $| \hat{L}_{\mathrm{Geo}} - L^*_{\mathrm{Geo}} |=2.7$ indicates that the proposed index terms miss more than two geographic names. 
Such reasoning is not covered by ordinary statistical text categorization methods. 
In addition, basic reliability indicators are included as features, such as content length (\#\_Char), individual term indicators (e.g.: TERM$_\mathrm{analysis}$), the number of out-of-vocabulary terms (\#\_W\_OOV), or different types of aggregations ($\Pi$) of the confidence scores of the assigned concepts. 
Finally, quality aspects are estimated using regression models based on the aforementioned features.

Development of the feature groups (Fig.~\ref{fig:qepred}: V, C, $\Pi$, LC) was driven by conceptual considerations. In particular, we wanted the features to represent:
\emph{imprecise input} (e.g., ``three European countries'': inherent ambiguity),
\emph{lack of input information} (e.g., title with fewer than 4 words: information is scarce),
as well as
\emph{lack of knowledge} (e.g., ``On Expected Effects of Brexit on European law'': information is present but can not be interpreted, if the term ``Brexit'' has not been observed before).

In general, the architecture of Qualle is a framework which, for example, allows to apply arbitrary regression methods for quality estimation.
Since the number of completely correct records in automatic subject indexing is extremely low,
we do not consider re-ranking by MaxEnt, which has been investigated for record-based confidence estimation in information extraction~\cite{Culotta2004}.
%
%
In this paper, we focus our analysis of the deep quality estimation regression approach on document-level recall. 
In addition, basic indicators have been considered for document-level precision estimation, that is, the mean ($\Pi_{\mathrm{mean}}$), product, median and minimum of the confidence values of the assigned concepts.

\section{Experiments}
\label{sec:experiments}
The experiments are centered around the following questions:
\begin{enumerate}[topsep=1pt, partopsep=1pt, leftmargin=25pt]
  \item[Q1:] Do predictions of recall and actual recall correlate with each other? 
  \item[Q2:] How accurate are the recall estimates?
  \item[Q3:] Which of the feature groups  contribute most to recall prediction? 
  \item[Q4:] What are the effects of filtering based on recall estimates on coverage and document-level precision? 
\end{enumerate}
which are relevant in practice.
Ranking documents by document-level recall (Q1) allows to separate high-recall documents from low-recall documents. 
Accurate estimates (Q2) allow to control filtering with meaningful constraints.
Applicability of the filtering approach would, however, be prevented if either document-level precision was decreased considerably or if the number of documents passing the filter was too low (Q4).

\subsection{Setup}
We evaluate the approach in two domains. 
We first perform a basic experiment on legal texts, addressing questions Q1 and Q2. 
Subsequently, 
we go into details regarding economics literature,
treating questions Q1-Q4.
The second experiments use more, and more elaborated features that notably exploit the categorisation of concepts into a hierarchy, as described in Section~\ref{sec:qe:qualle}.

The adequacy of quality estimation is measured in two ways.
Since perfect quality estimates follow their corresponding actual counterparts linearly, 
we consider the Pearson product-moment \emph{correlation} coefficient $\rho$ for (Q1)\footnote{If only ranking is relevant, rank-based correlation coefficients should be considered.}.
A strong correlation between predicted and true quality allows to order documents correctly,
that is, corresponding to the true performance. 
$\rho$ has been used in related studies~\cite{Culotta2004}.
For measuring the exactness of estimated recall values (Q2),
we consider the \emph{mean squared error} (MSE).
To gain knowledge about the utility of the feature groups (Q3),
we perform a systematic analysis of different configurations.
Feature groups are removed separately from the complete set of features (ablation study),
and measurements are also collected for each feature group alone (isolation study). 
%
Question Q4 was addressed by evaluating different thresholds on estimated recall 
and measuring average true precision and recall over the corresponding selected documents. In addition, the \emph{coverage}~$= \frac{ |\left\{ \mathcal{D}_{\mathrm{selected}} \right\}| }{N} $ was measured, with $N$ being the total number of documents and $\mathcal{D}_{\mathrm{selected}}$ the selected subset of the whole data set. We also 
report the relative recall \emph{gain} (RG) on theses subsets. 
The accuracy of initial multi-label classification is reported briefly for comparability, using metrics as described in Sect.~\ref{sec:qe}. 


Regarding law, we employ the EURLEX~\cite{LozaMencia2010} dataset to address Q1 and Q2.  
It comprises 19,314 documents, each having 5.31 EUROVOC\footnote{\url{http://eurovoc.europa.eu/}, accessed: 31.12.2017} subject terms on average. 
For further details on the data set, refer to~\cite{LozaMencia2010} and the website of the dataset\footnote{\url{http://www.ke.tu-darmstadt.de/resources/eurlex}, accessed 31.12.2017}.
Please note that the experiments in this paper only use the titles rather than the full text of the documents and that different train/test splits were used.
Regarding economics, we use three datasets,
which comprise roughly 20,000 (\Dkw), 60,000 (\Df), and 400,000 documents (\Dxl), respectively.
Each document is associated with several descriptors, for instance $5.89$ on average for \Dxl, from the STW Thesaurus for Economics (STW)\footnote{\url{http://zbw.eu/stw/version/latest/about.en.html}, accessed: 09.01.2018}.
Both, the STW and EUROVOC, comprise thousands of concepts, yielding challenging multi-label classification tasks.

For each data set, we perform cross validation with 5-folds.
And for each of those 5 runs,
we apply nested cross validation runs, likewise with 5 folds used for parameter optimization and learning of quality estimation,
That is, each training set is subdivided into \emph{dev-train} and \emph{dev-test} splits.
For validation, a new model is trained from random samples of the same size as one of the dev-train splits. As a consequence, the training and prediction of the classifier for label prediction as well as for the regressor for label calibration are carried out $5\cdot 6 = 30$ times for each collection. 
Quality estimation is evaluated on the corresponding \emph{eval-test} data folds.


For \emph{multi-label text classification}, we chose binary relevance logistic regression (BRLR) optimized with stochastic gradient descent (cf.~\cite{Wilbur2014,Bennett2017}).

Regarding \emph{reliability indicator variables,}
the EURLEX study relies on just two features:
the estimated number of concepts for the document, 
and the difference to the actually predicted number of concepts for the document by BRLR.
For the detailed study on economics documents,
all feature groups were employed (Sect.~\ref{sec:qe:qualle}).
%
Label calibration has been realized with tree-based methods
(EURLEX: ExtraTreesRegressor~\cite{Geurts2006}, Economics: GradientBoostingRegressor~\cite{Friedman2002}).
Only the total number of concepts per document is considered for EURLEX.
The economics experiments compute label calibration estimates
for the seven top categories of the corresponding thesaurus. 
For EURLEX and economics,
\#\_Char, \#\_WS and TERM$_i$ 
have been used as features for label calibration.

Several regression methods implemented in scikit-learn~\cite{Pedregosa2011} were considered for quality estimation.
%
%
For the EURLEX experiments, rather basic models like
LinearRegression and 
DecisionTreeRegression 
are tested, as well as ensemble machine learning methods, namely,
ExtraTrees~\cite{Geurts2006}, GradientBoosting~\cite{Friedman2002}, 
and AdaBoostRegressor~\cite{Drucker1997}.
different \emph{regression methods} were applied for recall estimation.
Regarding the more detailed experiments on economics research literature,
only the two regression models that performed best on EURLEX were investigated.
Extensive grid searches over various parameters of the models are left for future work.  

\subsection{Results}

\subsubsection{EURLEX}
From the different regression models,
LinearRegression produced the lowest correlation coefficient ($\rho = .214 \pm .026$) between predicted recall and true recall.
AdaBoostRegressor reached the highest correlation coefficient ($\rho = .590 \pm .013$) and the lowest mean squared error (MSE $=0.067 \pm 0.002$).
Only AdaBoostRegressor and GradientBoostingRegressor achieved correlation coefficients greater than $.500$. 
Although being worse than the AdaBoostRegressor on average,
the results for the ExtraTreesRegressor 
were more balanced.

\subsubsection{Economics}

Comparing the two selected regression methods, we found that the best configurations of GradientBoosting dominated the best configurations of AdaBoost on all datasets and with respect to both metrics ($\rho$, MSE).
Thus, Adaboost has been excluded from further analysis.

Table~\ref{tbl:exp:economics:ablation} offers the numbers for ablation and isolation of feature groups. 
For each collection, the complete set of features (first row corresponding to each collection) is always among the top configurations, where differences are not greater than the sum of their standard deviations. 
%
For all collections, the largest decrease in performance is recognized when the group of features related to label calibration is removed. 
In accordance, this feature group yields the strongest individual results. 
On \Dkw, its performance is close to the complete set of features.
For the collections with more data, the difference is more clear. 
%
%
Volume features, including length of the document, was found to be the lowest ranking group and has little impact when removed from the complete set of features.
In nearly all cases of configurations, 
more data yields higher correlation coefficients, however, not necessarily lower mean squared error.
In the following, we focus on reporting results regarding \Dxl. Figures for \Dkw\ and \Df\ were similar.

\begin{table*}[h!]
\caption{
Feature analysis for economics with GradientBoosting. $\checkmark$: presence of feature group.
$\Delta$: Difference in relation to complete set of features. 
\textdagger: Absolute difference to condition with all features is greater than the sum of their $sd$.
}
\centering
\begin{tabular}{ll*{4}{p{1.5em}}*{2}{r@{$\pm$}rr@{}r}}
\toprule
\multicolumn{2}{l}{Configuration} & 
V & C & LC & $\Pi$ & 
$\rho\;$ & std & $\Delta_\rho$ & &
MSE$\;$ & std & $\Delta_{\mathrm{MSE}}$ & \\ 
  \midrule
  \Dkw &  & \checkmark & \checkmark & \checkmark & \checkmark & 0.597 & 0.014 & -0.0\% &  & 0.039 & 0.001 & -0.0\% &  \\[2pt]
  \Dkw & \multirow{4}*{\rotatebox{90}{ablation}} & & \checkmark & \checkmark & \checkmark & 0.596 & 0.014 & -0.2\% &  & 0.040 & 0.001 & 0.2\% &  \\ 
  \Dkw &  & \checkmark &   & \checkmark & \checkmark & 0.595 & 0.015 & -0.3\% &  & 0.039 & 0.001 & -0.6\% &  \\ 
  \Dkw &  & \checkmark & \checkmark & \checkmark &   & 0.583 & 0.015 & -2.3\% &  & 0.040 & 0.001 & 1.8\% &  \\ 
  \Dkw &  & \checkmark & \checkmark &   & \checkmark & 0.384 & 0.005 & -35.6\% & \textsuperscript{\textdagger} & 0.050 & 0.001 & 26.5\% & \textsuperscript{\textdagger} \\[2pt]
  \Dkw & \multirow{4}*{\rotatebox{90}{isolation}} &   &   & \checkmark &   & 0.569 & 0.014 & -4.7\% & \textsuperscript{\textdagger} & 0.041 & 0.001 & 2.6\% &  \\ 
  \Dkw &  &   &   &   & \checkmark & 0.362 & 0.007 & -39.3\% & \textsuperscript{\textdagger} & 0.051 & 0.001 & 28.0\% & \textsuperscript{\textdagger} \\ 
  \Dkw &  &   & \checkmark &   &   & 0.196 & 0.013 & -67.1\% & \textsuperscript{\textdagger} & 0.056 & 0.001 & 41.1\% & \textsuperscript{\textdagger} \\ 
  \Dkw &  & \checkmark &   &   &   & 0.128 & 0.008 & -78.6\% & \textsuperscript{\textdagger} & 0.056 & 0.001 & 43.0\% & \textsuperscript{\textdagger} \\ 
  
  \midrule 
  \Df &  & \checkmark & \checkmark & \checkmark & \checkmark & 0.617 & 0.011 & -0.0\% &  & 0.043 & 0.000 & -0.0\% &  \\[2pt] 
  \Df & \multirow{4}*{\rotatebox{90}{ablation}} &   & \checkmark & \checkmark & \checkmark & 0.615 & 0.010 & -0.3\% &  & 0.044 & 0.000 & 0.3\% &  \\ 
  \Df &  & \checkmark &   & \checkmark & \checkmark & 0.602 & 0.009 & -2.5\% &  & 0.044 & 0.001 & 1.8\% &  \\ 
  \Df &  & \checkmark & \checkmark & \checkmark &   & 0.600 & 0.010 & -2.8\% &  & 0.044 & 0.000 & 2.4\% & \textsuperscript{\textdagger} \\ 
  \Df &  & \checkmark & \checkmark &   & \checkmark & 0.420 & 0.009 & -31.9\% & \textsuperscript{\textdagger} & 0.055 & 0.001 & 26.1\% & \textsuperscript{\textdagger} \\[2pt] 
  \Df & \multirow{4}*{\rotatebox{90}{isolation}} &   &   & \checkmark &   & 0.574 & 0.005 & -6.9\% & \textsuperscript{\textdagger} & 0.046 & 0.001 & 5.4\% & \textsuperscript{\textdagger} \\ 
  \Df &  &   &   &   & \checkmark & 0.391 & 0.011 & -36.6\% & \textsuperscript{\textdagger} & 0.056 & 0.001 & 28.7\% & \textsuperscript{\textdagger} \\ 
  \Df &  &   & \checkmark &   &   & 0.216 & 0.017 & -64.9\% & \textsuperscript{\textdagger} & 0.062 & 0.001 & 43.9\% & \textsuperscript{\textdagger} \\ 
  \Df &  & \checkmark &   &   &   & 0.069 & 0.009 & -88.8\% & \textsuperscript{\textdagger} & 0.064 & 0.001 & 48.2\% & \textsuperscript{\textdagger} \\ 
  
  \midrule 
  \Dxl &  & \checkmark & \checkmark & \checkmark & \checkmark & 0.648 & 0.002 & -0.0\% &  & 0.042 & 0.000 & -0.0\% &  \\[2pt]
  \Dxl & \multirow{4}*{\rotatebox{90}{ablation}} & \checkmark & \checkmark & \checkmark &   & 0.649 & 0.001 & 0.1\% &  & 0.042 & 0.000 & -0.1\% &  \\ 
  \Dxl &  &   & \checkmark & \checkmark & \checkmark & 0.648 & 0.001 & 0.0\% &  & 0.042 & 0.000 & 0.2\% &  \\ 
  \Dxl &  & \checkmark &   & \checkmark & \checkmark & 0.644 & 0.002 & -0.6\% & \textsuperscript{\textdagger} & 0.042 & 0.000 & 0.8\% & \textsuperscript{\textdagger} \\ 
  \Dxl &  & \checkmark & \checkmark &   & \checkmark & 0.528 & 0.002 & -18.5\% & \textsuperscript{\textdagger} & 0.050 & 0.000 & 19.5\% & \textsuperscript{\textdagger} \\[2pt]
  \Dxl & \multirow{4}*{\rotatebox{90}{isolation}} &   &   & \checkmark &   & 0.640 & 0.001 & -1.3\% & \textsuperscript{\textdagger} & 0.043 & 0.000 & 1.1\% & \textsuperscript{\textdagger} \\ 
  \Dxl &  &   &   &   & \checkmark & 0.511 & 0.002 & -21.2\% & \textsuperscript{\textdagger} & 0.051 & 0.000 & 21.6\% & \textsuperscript{\textdagger} \\ 
  \Dxl &  &   & \checkmark &   &   & 0.225 & 0.003 & -65.3\% & \textsuperscript{\textdagger} & 0.064 & 0.000 & 51.6\% & \textsuperscript{\textdagger} \\ 
  \Dxl &  & \checkmark &   &   &   & 0.122 & 0.002 & -81.1\% & \textsuperscript{\textdagger} & 0.065 & 0.000 & 55.3\% & \textsuperscript{\textdagger} \\ 

 \bottomrule
\end{tabular}
\label{tbl:exp:economics:ablation}
\end{table*}

\begin{figure}[t]
\begin{tabular}{cc}
 \includegraphics[width=.49\linewidth,align=c]{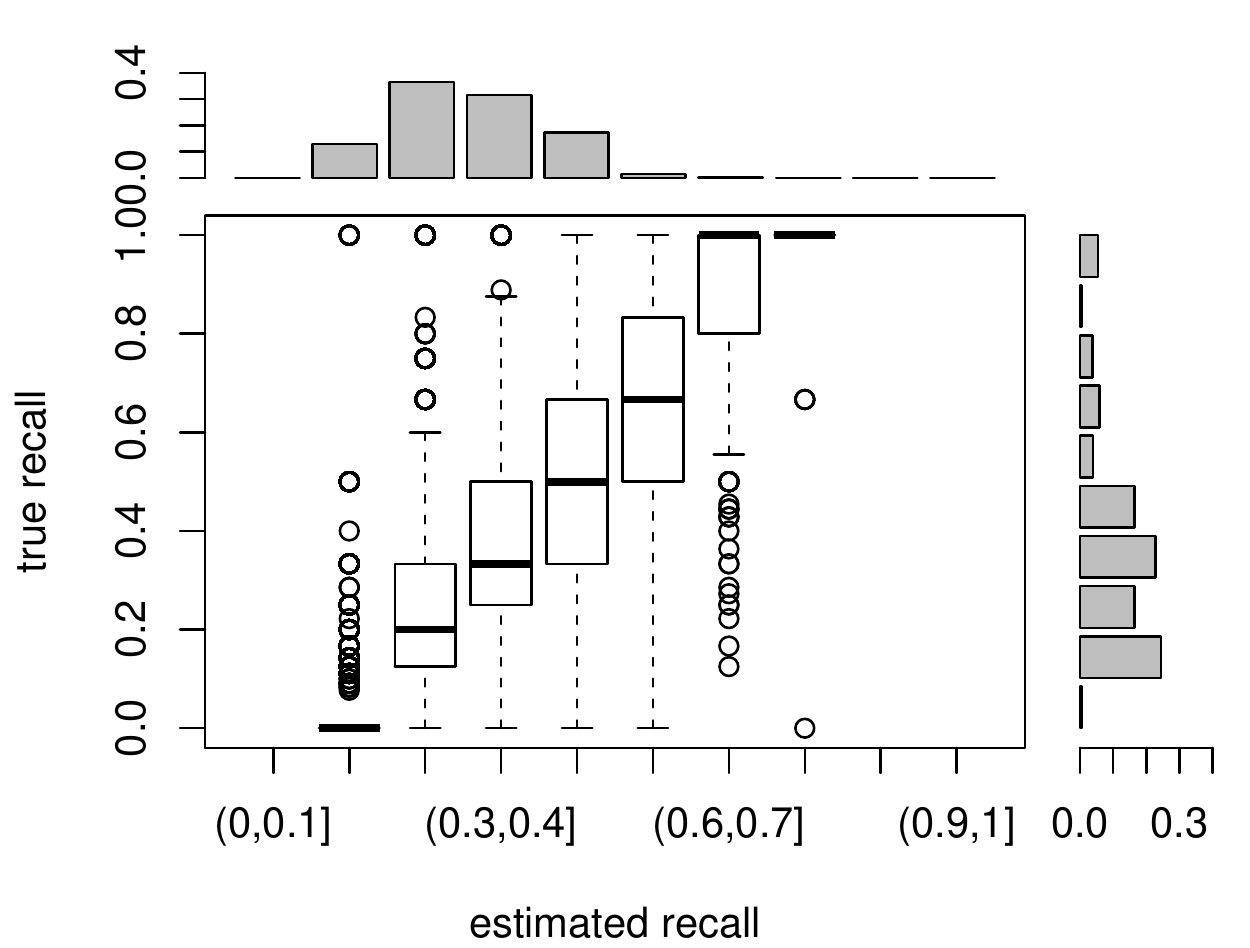} &
\includegraphics[width=.49\linewidth,align=c]{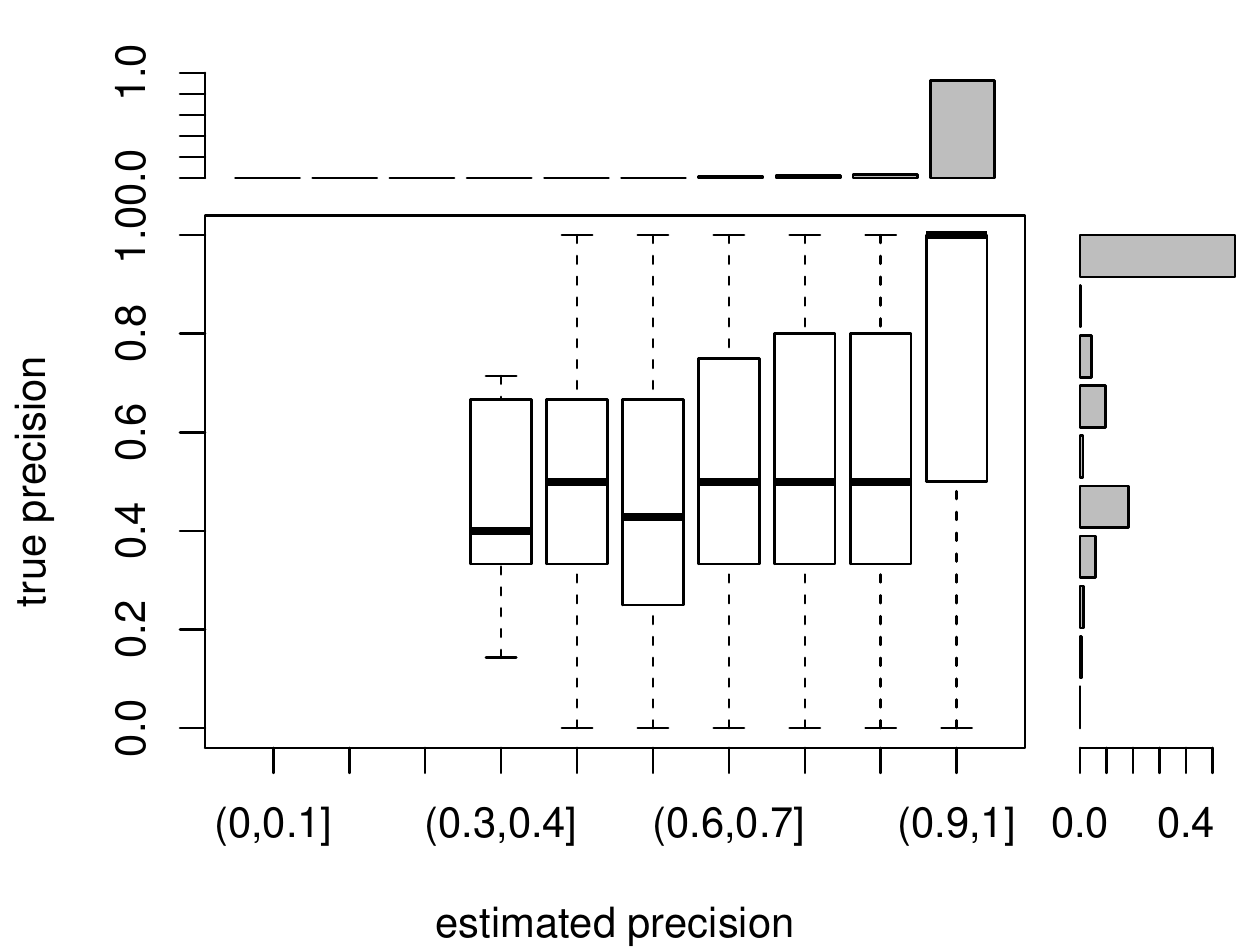}\\
a) & b)
\end{tabular}
\caption{
Quality estimates and true values (Economics: \Dxl). a) Recall estimation by Qualle, b) Precision score by product of concept confidence values.
Marginal distributions (bin count/total count) are shown on the top and on the right, respectively.
}
\label{fig:exp:economics:precrec:truepred}
\end{figure}

Figure~\ref{fig:exp:economics:precrec:truepred}a) depicts recall estimation results for \Dxl. 
The plot illustrates the degree of linear relation and also reveals the distributions of estimated and true recall values. Most of the documents have a true recall that is less than $60\%$.
Regarding the scoring functions for document-level precision,
the product of concept-level confidence scores exhibited the highest correlations for \Dkw\ and \Df, however, still staying below $.500$. 
On \Dxl, all scoring functions were very close to each other, and their correlation coefficients were above $.500$. Figure~\ref{fig:exp:economics:precrec:truepred}b) depicts results for the product of concept confidence values.

\begin{figure}[tbp]
\centering
\includegraphics[width=\linewidth,align=c]{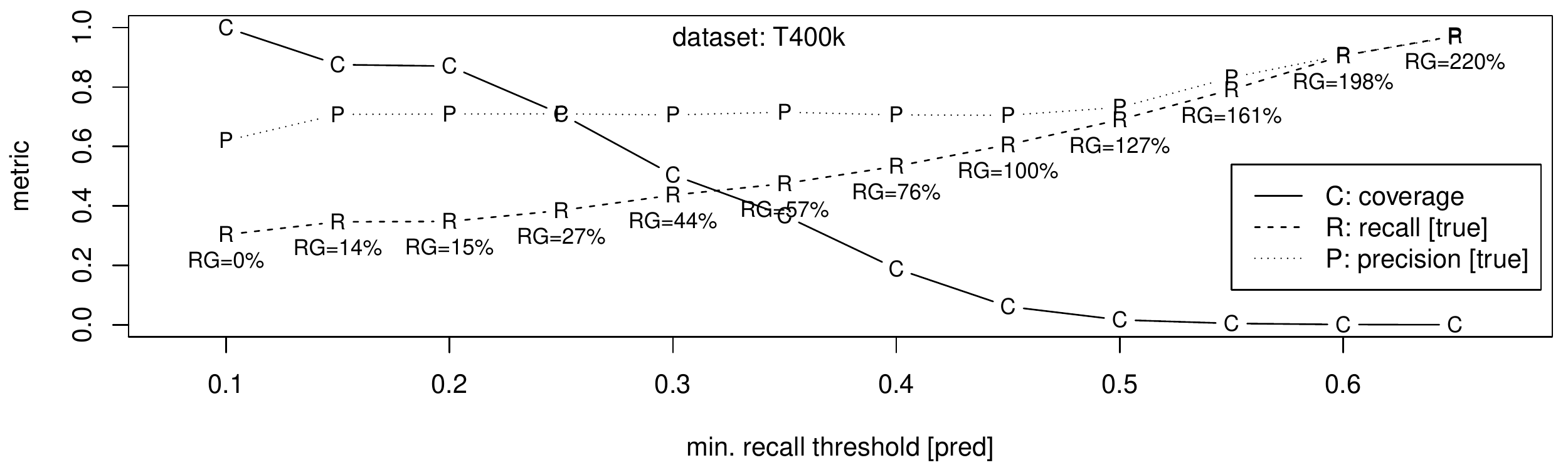}
\caption{
Quality vs. coverage (collection: \Dxl):
coverage, 
mean document-level true recall and true precision
for different predicted recall thresholds.
RG: relative gain in document-level recall on selected subset 
compared to 
the full dataset.
}
\label{fig:exp:economics:qualitycoverage}
\end{figure}

%
%
Finally, Fig.~\ref{fig:exp:economics:qualitycoverage} visualizes
how different thresholds on estimated recall affect 
properties of the resulting document selections.
The plot therefore shows coverage, as well as mean document-level true recall and true precision.
When constraining estimated recall to be at least 30\%,
a gain RG=44\% of true recall in relation to the measure on the comple collection could be achieved on \Dxl.
The most relevant message that can be drawn from Fig.~\ref{fig:exp:economics:qualitycoverage}  is that 
the precision on the selected subsets remained the same or even increased, when putting harder constraints on estimated recall.

%
\subsubsection{Multi-label Classification}
The performance of the BRLR approach was not in the focus of the study, yet BRLR turned out to be a reasonable choice. For instance, it reached sample-based average $f_1 = 0.361$, precision $= 0.528$, recall $= 0.327$ on \Dkw, and $f_1 = 49.1\%$ (micro avg.) on EURLEX. These figures broadly correspond to related studies.

\subsection{Discussion}

The basic set of features used on EURLEX 
reached respectable correlations ($\rho > .500$) between predicted and true recall (Q1) 
only for the sophisticated machine learning methods AdaBoost and GradientBoosting.
Differences in the balance of predictions 
should be considered for applications, just like the notable amount of variance that remains around the predicted values (Q2).
In summary,
the outcome of the EURLEX study suggested that recall estimates that are useful for filtering are feasible, motivating investigation of more complex configurations.


Looking at the outcome of the experiments on the economics datasets,
especially Fig.~\ref{fig:exp:economics:qualitycoverage},
our results show that the proposed quality estimation approach can be successfully applied to identify subsets of document collections where soft constraints on precision as well as recall are met (Q4). 
Finally, it remains a decision depending on the application context to make trade-offs according to multi-criteria objectives, which notably comprise coverage.
%
Regarding recall, ranking and accuracy of predictions are sufficient enough (Q1, Q2). 
Interestingly, precision was not affected negatively (cf. Fig.~\ref{fig:exp:economics:qualitycoverage}).
Based on Table~\ref{tbl:exp:economics:ablation},
applications should consider the full set of features, which belongs to the top performing configurations in all cases and outperformed individual feature groups. 
Label calibration information is found to be a strong individual predictor. It is the most relevant reliability indicator (Q3) compared to the volume, content, and concept-confidence related feature groups.
%
%
The mean squared errors of predictions indicate that a considerable amount of vagueness remains (Q2). Possibly, it may be caused by the errors in concept assignments, 
which influence the label calibration related features.


%
The experimental results highlight the inevitable difficulties (cf. Sect.~\ref{sec:qe:analysis}) in multi-label text classification, namely, 
suffering from low document-level recall when the model misses knowledge (either dictionary entries or training examples), or when the observed input is inherently ambiguous.
Quality estimation enables to handle such issues by controlling, that is, making trade-offs between quality and coverage.
Since the proposed approach is not bound to specific MLC 
or regression methods, further progress in this regard can be integrated and is assumed to improve collection coverage.
Another direction for future work is to consider alternative quality metrics that take semantic relations into account (see e.g.,~\cite{Medelyan2006,Neveol2006}).

\section{Conclusion}
\label{sec:conclusion}
In order to assure data quality in operative information retrieval systems with large and diverse datasets, we investigated an important yet less addressed research topic, namely quality estimation of automatic subject indexing with a focus on the document level.
Our experimental results on two domains spanning over collections of different sizes
show that the proposed multi-layer architecture is effective and thus enables to control quality in settings where high standards have to be met.
The approach allows to define different thresholds, which resulted in considerable gains of document-level recall, while upholding precision at the same time. 
Label calibration was the most relevant reliability indicator.

\bibliographystyle{splncs}
\bibliography{literature}


\end{document}